\documentclass{article}              

\usepackage{cite}
\usepackage{amsmath}                
\usepackage{amsfonts}               
\usepackage{algorithmic}            
\usepackage{array}
\usepackage{hyperref}
\usepackage{cancel}

\usepackage{fixltx2e}
\usepackage{graphicx}
\usepackage{url}

\newcommand{\refmmlist}[2]{\href{https://www.cantab.net/users/mmlist/#1}{#2}}
\newcommand{\refclass}[3]{\refmmlist{MML/A/doc/#1/#2.html}{#3}}
\newcommand{\refmember}[4]{\refmmlist{MML/A/doc/#1/#2.html\##3}{#4}}

\newcommand{\mytilde}{\raise.17ex\hbox{$\scriptstyle\mathtt{\sim}$}}

\newcommand{\ignore}[1]{}

\begin{document}
\title{Have Object-Oriented Languages Missed a Trick with Class Function and its Subclasses?}   

\author{Lloyd Allison, \\
  Faculty of Information Technology, Monash University, \\
  Clayton,
  Victoria 3800,
  Australia. \\
  lloyd.allison@monash.edu
}

\date{}

\maketitle   

\begin{abstract}
Compared to functions in mathematics,
functions in programming languages seem to be under class-ified.  
Functional programming languages based on the lambda calculus
famously treat functions as first-class values.
Object-oriented languages have adopted ``lambdas'', notably for
call-back routines in event-based programming.
Typically a programming language has functions,
a function has a type, and
some functions act on other functions and/or return functions but
there is generally a lack of
(i)\,``class Function'' in the OO sense of the word class
and particularly
(ii)\,subclasses of Function for functions having specific properties.
Some such classes are presented here and programmed in
some popular programming languages
as an experimental investigation into
OO languages missing this opportunity.
  \\ \\
\textit{keywords:} class Function,
  Differentiable, inverse, apply, compose
\end{abstract}

\section{Introduction}
\label{sec:Intro}

Computer programming languages can be compared in many different ways.
One important basis for comparison
is how languages treat functions (subroutines, procedures, methods).
This paper suggests that object-oriented languages,
and languages with object-oriented features,
would benefit from having a \textit{standard} \texttt{class Function}
with methods \texttt{apply} (\texttt{()}) and
\texttt{compose} (\texttt{$\circ$}) and,
in addition to that, various subclasses of \texttt{class Function}
for functions having extra properties.
Every function would be an object, an instance of
\texttt{class Function}\footnote{\textit{function} is
  a keyword in many programming languages so merely to avoid confusion
  \textit{funktion} and \textit{Funktion} are sometimes used in what
  follows to distinguish something new being suggested,
  implemented or added.}
or some subclass of it.

\begin{figure*}
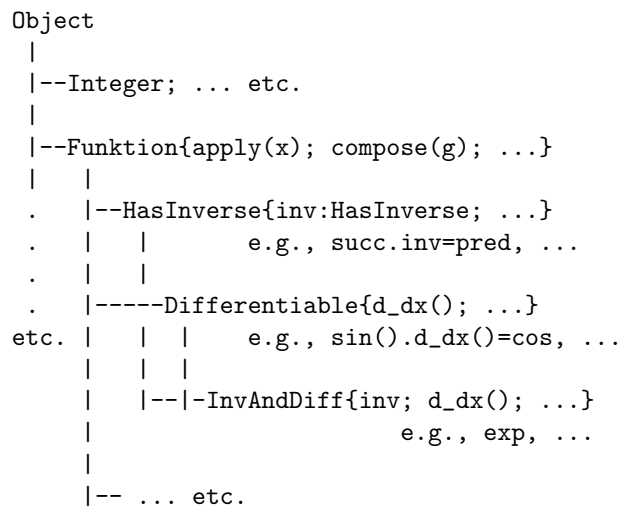

\begin{verbatim}
  Object
   |
   |--Integer; ... etc.
   |
   |--Funktion{apply(x); compose(g); ...}
   |   |
   .   |--HasInverse{inv:HasInverse; ...}
   .   |   |       e.g., succ.inv=pred, ...
   .   |   |
   .   |-----Differentiable{d_dx(); ...}
  etc. |   |  |    e.g., sin().d_dx()=cos, ...
       |   |  |
       |   |--|-InvAndDiff{inv; d_dx(); ...}
       |                      e.g., exp, ...
       |
       |-- ... etc.
\end{verbatim}
\caption{Some Possible Funktion SubClasses}
\label{fig:Funktion}
\end{figure*}

As just a first brief examination of possible subclasses
(figure \ref{fig:Funktion}) we consider classes of functions
that have an inverse such as
\texttt{succ} (successor) and \texttt{pred} (predecessor)
and
of functions that have a derivative such as
\texttt{sin} (sine) and \texttt{cos} (cosine),
and
functions that have both an inverse and a derivative
such as \texttt{exp} and \texttt{log}.
Of course inverses and derivatives are themselves functions.
(Note that in general a derivative must be created lazily,    
on demand, because the derivative may well have a derivative, and so on,
possibly infinitely -- consider $f(x)=e^{2x}$.)

\textit{Orthogonality} in the context of a programming language is where
  ``The number of independent primitive concepts has been minimized in
  order that the language be easy to describe, to learn, and to implement.
  On the other hand, these concepts have been applied ``orthogonally'' in
  order to maximize the expressive power of the language while trying to
  avoid deleterious superfluities''
  \cite{A68} (sec.0.1.2, p.9) --
it is a condition on the ``dimensions'' of the language that,
as far as possible, dimensions should be independent of each other.
For example,
one dimension might be the \textit{types}\footnote{In the
  interests of uniformity, liberties will sometimes be taken
  with the names of certain types and other terminology which
  may vary from language to language, for example,
  \textbf{Z} v.  int v. integer,
  \textbf{R} v. real v. float,
  $(t \rightarrow u)$ v. proc(t)u v. u\;function(t),
  abstract v. virtual, and so on.}
of values that can be computed
(int, real, bool, char, structure, array, function, ...).
A second dimension might be
the \textit{context} in which some value can appear
(the RHS of a declaration of a named constant,
the RHS of an assignment to a variable,
an element of a structure,
an element of an array,
an actual parameter of a function,
the result of a function, ...),
then if some combination of choices from
two or more independent dimensions is meaningful --
such as \texttt{[sqrt, sin, cos, aUsersFn]} that is an
\texttt{array} of \texttt{functions} from \texttt{real} to \texttt{real} --
then it is a good thing for it to be allowed in the language \cite{Str67}.
Well, in object oriented languages, if
\texttt{77} can be an instance of \texttt{class Integer} and
``\texttt{hello world}'' can be an instance of \texttt{class String},
cannot \texttt{pred} be an instance of \texttt{class Function}?

Functional programming languages
and in particular Haskell \cite{Haskell98, Haskell2010}
are often held up as orthogonal programming languages.
Haskell does have a rather elegant
distinction between value, type and (type-)
class:\footnote{There are clear differences between
  type-classes and OO classes but there are also clear overlaps
  and this paper is not about whether or not Haskell is object-oriented.
  Haskell also uses ``instance'' differently.}
a value has (or is a member of) a type and
a type can be an instance of one or more \textit{type classes},
for example, \texttt{7} has the type \texttt{Int} and
\texttt{Int} is an instance
of the \texttt{Eq} (equality, \texttt{==}) type class and
of the \texttt{Num} (numeric) type class.
In fact Haskell has no less than nine numerical classes
(\texttt{Num}, \texttt{Bounded}, \texttt{Enum}, \texttt{Real},
\texttt{Fractional}, \texttt{Integral},
\texttt{RealFrac}, \texttt{Floating}, \texttt{RealFloat}) \cite{HaskellPredef}
but none for functions.

Those functional programming languages
that are based on the lambda calculus \cite{Chu41},
arguably have the high ground when it comes to dealing with functions,
holding to ``functions are first class values''
in that functions can appear in data structures and can be
passed to and returned by functions --
consider \texttt{map}, \texttt{reduce} and so on.
However Haskell, for example, does not have a type class \texttt{Function}
having methods apply and compose,
although opinion \cite{H2002} seems to be that it would be possible.
Note in passing that Haskell has
an explicit visible representation of apply, \texttt{f\$x},
in addition to the usual implicit apply, \texttt{f\,x},
also that as \texttt{(x)=x} so \texttt{f\,x=f(x)} --
the usual application operator is implicit, invisible.

There are some other
functional programming languages, not strongly based on lambda calculus,
notably APL \cite{APL} and its descendant J \cite{J}
that have a rich collection of functions and operators acting
on functions and leading to famously compact programs for many problems.
Backus \cite{Bac78} in his later work
advocated this algebraic style of programming language:
``Most programs written today are
``object-level''\footnote{Note, this before
         the growth of object-oriented programming, OOP.}
programs.
That is, programs describe how to combine various ``objects''
(i.e., numbers, symbols, arrays, etc) to form other objects until
the final ``result objects'' have been formed. ... Lambda calculus based
languages, such as LISP and ISWIM [Landin 66]\cite{Lan66}, are also,
in practice, object level languages, although they have the means
to be more.'' \cite{Bac81}.

Looking at mathematics, functions are used and
are themselves operated on in many ways --
$f\,x = f(x), f^{-1}$ the inverse of a suitable $f$,
$argmin\,f$, $argmax\,f$,
$f \circ g$ the composition of $f$ and $g$,
$f^2 = f \circ f$,
$f+g$, $f-g$, $f \times g$,
$f'$ the derivative of $f$,
$\int_{lo}^{hi}f(x)\,dx$ the integral of $f$.
Familiar equations include
$(f+g)' = f'+g'$,
$(f \circ g)^{-1} = g^{-1} \circ f^{-1}$, and
the chain rule $(f \circ g)' = (f' \circ g) \times g'$;
$f+g$, $f \circ g$ and so on are differentiable if $f$ and $g$ are, and
$f \circ g$ has an inverse if $f$ and $g$ do.
It is possible to find lists of many more operators on functions,
for example see \cite{operators}.

The main aim of this paper is to examine just \textit{some} of
the kinds of classes and subclasses of function that it would
be nice to have in a programming language
to support this kind of algebraic programming.
A further aim is to see how
close some popular programming languages are
to providing this algebraic style as standard or
to allowing it to be programmed;
they can all do the latter more or less easily
but more interesting is how close are they to the former?
There follow an example application and then
some trials in Javascript, Java, C++, Python and,
without entering the argument about whether or not it is object-oriented,
Haskell.

\section{A Use}
\label{sec:Application}

An early motivation for considering subclasses of class Function
was the transformation of statistical models \cite{All22}
in a package of statistical and inductive inference software \cite{All18}
for machine learning problems.
The central classes in that package are
\refclass{mml}{UPModel}{\texttt{UPModel}} (unparameterised models),
\refclass{mml}{Model}{\texttt{Model}} (parameterised models) and
\refclass{mml}{Estimator}{\texttt{Estimators}} and
there are many instances of these including
the \refmember{mml}{MML}{Normal}{Normal} probability distribution and
\refmember{mml}{MML}{N01}{N01} ($N_{0,1}$).

As an example of a transformed model,
a log-Normal probability distribution over a data-set of
positive continuous data, \texttt{ds},
can be estimated by fitting a Normal distribution to
the \texttt{log}s of the data, to \texttt{map log ds}.
For continuous data there is also the matter of the
accuracy of measurement (AoM) of the data
which the inference package requires,
and this is different in the transformed data
from the raw data \texttt{ds} by a factor of \texttt{log'}.
Another thing that a parameterised Model
should be able to do is to generate a random sample of data;
in the case of a log-Normal generate a sample from the underlying Normal
and apply the inverse of \texttt{log}, that is \texttt{exp},
to the members of that sample.
So in general,
if a model of continuous (Real, \textbf{R}) data
is transformed by a function $f : \textbf{R} \rightarrow \textbf{R}$
then the transformed model
needs access to $f^{-1}$ and to $f'$.
Thus, a constructor for a
\refclass{mml}{UPModel.Transform}{transformed} model
takes a function \texttt{f} that must be in the
\refclass{la/la}{Function.Cts2Cts.WithInverse}{subclass} of \texttt{Function}
for functions that have an inverse and a derivative.

The situation for multivariate continuous data
($\textbf{R}^D$) and models of such data transformed by a function
$f : \textbf{R}^D \rightarrow \textbf{R}^D$
is similar to the above except that
the transformed model needs the Jacobian of $f$
rather than a single derivative and
there is a subclass for such functions.
A concrete case involves cartesian and polar data and
mutual inverse functions $polar2cartesian$ and $cartesian2polar$
and their Jacobians.
%

\begin{figure*}
\fontsize{10}{10}
\begin{verbatim}
var NN = 0;  // just a counter, f0, f1, ...
class Funktion
 { constructor(f)
    { this.apply = f; this.NN=NN; NN++; this.hasInv=false; }
   apply = function(x) { error("apply undefined as yet"); }
   compose = function(G)
    { const F  = this;
      const FG =
         new Funktion( function(x){ return F.apply(G.apply(x)); } );
      FG.toString = function() { return "("+F+"."+G+")"; };
      return FG;
    }
   plus = function(G){ ... }  // (f+g)(x)=f(x)+g(x)
   toString = function() { return "f"+this.NN; }
   ...
 }

class HasInverse extends Funktion // subclass of Funktion
 { constructor(f, finv) //NB f a function, finv a Funktion
    { super(f); this.inv=finv; this.hasInv=true; }
   compose = function(G)
    { ...NB. (f o g).inv = (g.inv)o(f.inv) if g has an inverse...
   ...
 }

class Differentiable extends Funktion
 { constructor(f, make_d_dx)
    { super(f); if(make_dx!=null) this.make_d_dx=make_d_dx; }
   cache_d_dx = null;
   make_d_dx = function()
    {...} //default implementation by finite differences
   d_dx = function()
    { if( this.cache_d_dx == null )
         this.cache_d_dx = this.make_d_dx(); // lazy!
      return this.cache_d_dx;
    }
   compose = function(G) { ...} // differentiable if G is
   plus = function(G) { ... }   //     --"--
   ...
 }

class InvAndDiff extends Differentiable
 { constructor(f, derivative)
    { super(f, derivative);
      this.hasInv = true;
      this.inv = null; // must define it eventually
    }
   ...
 }
\end{verbatim}
\caption{Possible class Funktion in Javascript}
\label{fig:Javascript}
\end{figure*}

\section{Trials}
\label{sec:Trials}

This section looks at implementing classes, and instances,
of functions that may or may not
(i)~have an inverse,
(ii)~be differentiable, and
(iii)~both have an inverse and be differentiable,
as in figure \ref{fig:Funktion}.
This is by no means all the subclasses of Function that one might like
to have in general programming but it provides a simple fixed target
to see what one can do by way of classes of function in
the object-oriented languages Javascript, Java, C++ and Python,
and also Haskell,
as they stand.
Defining other operations on functions is clearly possible and, for example,
defining integrals would have a lot in common with defining derivatives.
Some other possibilities are discussed in section \ref{sec:Subclasses}.

The programming languages are chosen simply on the grounds
that they are popular;
this is not language wars and there is no claim that
they are the most interesting, most important or
best object-oriented languages.
It is no surprise that
each of the languages (and doubtless many others)
is able to define classes and objects that contain
the required information for functions
that may have inverses, derivatives and the like but
the questions are
(i)\,how convenient is each case and
(ii)\,why didn't some of the languages themselves
take more advantage of the possibilities?

\subsection{Javascript}
\label{sec:Javascript}

Javascript \cite{JS} functions are first-class values
in that they can be
named or anonymous, passed as parameters, returned as results,
stored in data-structures and so on.
One can define a new class \texttt{Funktion} (figure \ref{fig:Javascript})
and subclasses, such as \texttt{HasInverse},
but one has to write \texttt{f.apply(x)} rather than \texttt{f(x)}
to call a Funktion \texttt{f} as opposed to a
normal (Javascript) \texttt{function}.
Note \texttt{cache\_d\_dx}, \texttt{make\_d\_dx()} and
\texttt{d\_dx()} making the derivative only on demand.
The default implementation of derivatives can be by finite differences but
this can be overridden if there is a closed form such as $sin' = cos$.

Javascript does not support multiple inheritance
so \texttt{InvAndDiff} can only be a subclass of one of \texttt{Inverse}
or \texttt{Differentiable} but using a boolean flag, \texttt{hasInv},
in \textit{all} \texttt{Funktion} classes
to indicate which have inverse is one possible work-around of many.

Given the above one can then, for example, make
\texttt{succ.inv} to be \texttt{pred} and vice versa,
\texttt{sin.d\_dx()} to be \texttt{cos},
\texttt{exp.d\_dx()} to be \texttt{exp} and
\texttt{exp.inv} to be \texttt{log}.

\subsection{Java}
\label{sec:Java}

\begin{figure*}
\begin{verbatim}
public static abstract class Funktion<T,U>  // T -> U
 { ...
   public abstract U apply(T x); // f(x)

   public <S> Funktion<S,U> compose(Funktion<S,T> g) // f o g
    { final Funktion<T,U> f = this;
      return new Funktion<S,U>()
       { public U apply(S x)
          { return f.apply(g.apply(x)); }; // f(g(x))
       };
    }

   public static class HasInverse<T,U> extends Funktion<T,U>
    { public HasInverse<U,T> inv; ...}

   public static class Differentiable extends Funktion<Double,Double>
    { protected Differentiable cache_d_dx = null;
      protected public void make_d_dx() {...set cache_d_dx...}
      public Differentiable d_dx()
       { if( cache_d_dx == null ) make_d_dx(); //lazy, by need
         return cache_d_dx;
       }
      public Differentiable compose(Differentiable g) {...}
    }

   public static InvAndDiff exp = new InvAndDiff()
    { public InvAndDiff inv ...}
 }
\end{verbatim}
\caption{Possible class Funktion in Java}
\label{fig:Java}
\end{figure*}

Figure \ref{fig:Java} shows a \texttt{class Funktion} in Java \cite{Java}.
Here, the options of
using type variables such as \texttt{T} and \texttt{U}, and of
making \texttt{HasInverse} and other subclasses of \texttt{Funktion}
inner classes have been taken.
Naturally the inverse of \texttt{f}, \texttt{f.inv},
itself has an inverse:~\texttt{f}.
Like Javascript,
Java does not allow multiple inheritance from full classes
so \texttt{InvAndDiff} cannot extend both \texttt{HasInverse} and
\texttt{Differentiable};
one might use a ``flag'' in \texttt{Funktion} or an interface
-- rather than \texttt{instanceof HasInverse} --
to indicate those subclasses (plural) of \texttt{Funktion} having inverses.

It is then possible to create new Funktion objects such as
\texttt{succ}, \texttt{pred}, \texttt{sin}, \texttt{exp}, and so on
but to use them we have to write \texttt{succ.apply(7)}
because the implicit apply, \texttt{()}, cannot be overridden.

Java classes \textit{similar} to the above were used
to transform statistical models \cite{All22} in a
package of statistical and inductive inference software \cite{All18}
for machine learning problems.

\subsection{C++} 
\label{sec:Cpp}

Figure \ref{fig:Cpp} shows a possible
\texttt{class Funktion} in C++ \cite{Cpp}.
It is often said the C++ functions cannot return general functions
because C++ does not fully support closures,
but this does not hold for user-defined class {Funktion};
note that \texttt{f.compose(g)}
is created in the heap (\texttt{return new Composition(this, g)})
and storage management will be an issue.

It is possible to overload the
implicit function application operator, (\texttt{()}),
so one can write \texttt{f(x)} to call a \texttt{Funktion}, \texttt{f}.

\begin{figure*}
\begin{verbatim}
template <class T, class U>
class Funktion  // T -> U
 { public:
    Funktion() { }
    std::string toString() { return "Funktion"; }

    template <class S>
    Funktion<S,U>* compose(Funktion<S,T>* g)
     { class Composition : public Funktion<S,U>
        { private:
             Funktion<T,U>* f; Funktion<S,T>* g;
          public:
             Composition(Funktion<T,U>* f, Funktion<S,T>* g)
              { this->f=f; this->g=g; }
             U apply(S n) { return f->apply(g->apply(n)); }
        };//Composition class
       return new Composition(this, g);  // i.e., in the heap
     };//compose(g)

    U operator()(T x) { return this->apply(x); };  // overload ()

    virtual U apply(T x) = 0;
 };//Funktion class
...

class : public Funktion<int,int>  // an example Funktion, sqr
 { public: int apply(int n) { return n*n; } } sqr;
\end{verbatim}
\caption{Possible class Funktion in C++}
\label{fig:Cpp}
\end{figure*}

\begin{figure*}
\begin{verbatim}
NN = 0  # global Funktion counter

class Funktion:
  def __init__(self, f):  # constructor
    self.apply = f
    self.toString = lambda : f"f{self.NN}"
    try:
      self.NN >= 0  # multiple inheritance means a constr. can be called x2
    except:         # self.NN does not yet exist, so ...
      global NN
      self.NN = NN  # each Funktion has a serial number
      NN += 1

  def __call__(self, x):  # make the Funktion callable
    return self.apply(x)

  def __str__(self):  # i.e., toString()
    return self.toString()

  def plus(self, g):  return lambda x: self(x)+g(x)
  ...

  def compose(self, g):  # self.compose(g)(x) = self(g(x))
    if not isinstance(g, Funktion):
      raise RuntimeError(f"{g} is not a Funktion")
    s_o_g = Funktion(lambda x : self(g(x)))
    s_o_g.toString = lambda : f"({self}.{g})"
    return s_o_g

class HasInverse(Funktion): ...
class Differentiable(Funktion): ...
class InvAndDiff(HasInverse, Differentiable): ...

\end{verbatim}
\caption{Possible class Funktion in Python}
\label{fig:Python}
\end{figure*}

\subsection{Python}
\label{sec:Python}

In Python \cite{Python} every value is
an instance of some class and
this has the ring of a nice clean design principle
but interestingly there are at least two classes of function --
class \path{builtin_function_or_method}  
for functions such as \texttt{abs} and \texttt{sin}
that come with the Python implementation, and
class \texttt{function} for \textit{user-defined} multi-line functions and
(anonymous) \texttt{lambda} expressions --
but both of these classes are \textit{direct} subclasses of \texttt{Object}
and they seem to be more about package management
than about classes as in having specific properties.

Note also that
it is not uncommon for languages in one programming paradigm to have
libraries of code to exploit programming techniques from another paradigm
and PyToolz \cite{pytoolz} is one such supporting
functional programming techniques in Python but
(it must be said, like Haskell)
PyToolz does not exploit subclasses of functions
in the ways discussed in this paper.

Instances of a Python class, \textit{any} class,
can be made callable, as in \texttt{f(x)},
if the class defines the method \texttt{\_\_call\_\_(self,x)}
-- which is itself an instance of class \texttt{method-wrapper}.
So \texttt{class Funktion} can be defined (figure \ref{fig:Python}) with
pretty much what we want syntactically.
If \texttt{f} is a Funktion one can therefore write \texttt{f(x)} to call it.
One does wonder if classes
\texttt{function} and \texttt{builtin\_function\_or\_method}
could be subclasses of a new class \texttt{Funktion}
(probably with a 'c' in the name rather than the 'k')
which could have the (abstract) method \texttt{\_\_call\_\_(self,x)}.
However,
taking Python as it is,
it is possible to, say, define class Funktion
with subclasses \texttt{HasInverse} and \texttt{Differentiable} and,
by multiple inheritance,
\texttt{InvAndDiff} for those Funktions
that have both an inverse and a derivative.
For example,
\texttt{succ.inv} is \texttt{pred} and vice versa,
\texttt{f.compose(g).inv} is
\path{g.inv.compose(f.inv)},  
and
\texttt{sin.d\_dx()} is \texttt{cos}
and so on.

However as things stand,
if one wants to bring \textit{built-in} functions, such as \texttt{sin},
into this hierarchy it is necessary to define \texttt{Funktion}
wrappers for them.

\subsection{Haskell}
\label{sec:Haskell}

In Haskell a value has a type and a type can be
an \textit{instance} of one of more type-classes.
There is a standard function type,
\texttt{t->u} (where \texttt{t} and \texttt{u} are type parameters), and
there is an explicit visible representation, \texttt{\$}, of application
as in \texttt{f\$x = f(x) = f\,x}.
There is no class Function
but it is possible to define a class \texttt{Funktion}
with 'methods' \texttt{apply}, \texttt{\$} and \texttt{compose}
if we \textit{hide} the language's standard uses of those names.
The function type \texttt{->} and
suitable user-defined types such as \texttt{Fn}
can be made instances of \texttt{Funktion}
(figure \ref{fig:Haskell}).

Subclasses of \texttt{Funktion} --
\texttt{HasInverse}     with method  \texttt{inv}, and
\texttt{Differentiable} with methods \texttt{d\_dx},
\texttt{d2\_dx2}, \texttt{plus}, \texttt{neg} and so on --
can also be defined.
Type \texttt{->} can be made an instance of \texttt{Differentiable}
by implementing \texttt{d\_dx} by finite differences.
If a closed form is known for a derivative
it can be provided as a second component of a suitable type,
for example,
\texttt{data FnD u v =
(Floating u, Floating v, u\mytilde{}v) => FnD (u->v) (FnD u v)}.
That is, the input and result types, \texttt{u} and \texttt{v},
of differentiable functions must be \texttt{Floating} types and
in fact must be the same (\texttt{\mytilde{}});
this is a use of Haskell's GADT (generalised abstract data types)
type extension.
One might fancy that \texttt{FnD} should instead have just one type
parameter rather than having two that must be the same
but if so its \textit{kind} would not be compatible with
class \texttt{Funktion}.

There is no need to create a common subclass of
both \texttt{HasInverse} and \texttt{Differentiable} --
a type \texttt{FnID} for such functions is simply
made an instance of both of them.

\vspace{3cm}  
The machinery above lets us define \texttt{Funktions} including the following:
\begin{verbatim}
  leng :: Fn String Int
  leng    = Fn (\str->length str)
  succ    = FnI (\n->n+1) pred  -- inverses
  pred    = FnI (\n->n-1) succ  -- inv pred = succ
  sine    = FnD sin cosine      -- differentiable
  cosine  = FnD cos (neg sine)  -- cosine' = -sine
  expe    = FnID exp loge expe  -- both inv and diff
  loge    = FnID log expe (diff2FnID oneOver)
  oneOver = FnID (\x->1/x) ... ...
\end{verbatim}
For example,
\texttt{succ} and \texttt{pred} are mutual inverses and
\texttt{sine} can be seen as a wrapper of \texttt{sin} from
the standard prelude.

There are some `wrinkles'.
If a function is differentiable then (in practice)
its derivative will be differentiable
but if it also has an inverse then
its derivative might or might not have an inverse.
Hence the \texttt{d\_dx} operator on such a \texttt{FnID} function
should either
always return a \texttt{FnD} or
return a \texttt{Maybe} value
but none of these are easily made
compatible with class \texttt{Differentiable};
for now the prototype code
sets the derivative's inverse to \texttt{undefined}.

Overall, Haskell enables most of what we want on the test-case classes
and it is hard to see why the like could not be provided as part of
the language and its standard library
in which case the \textit{standard} apply, \texttt(\$) and compose
could be methods of a standard class \texttt{Function} and so on.

\begin {figure*}
\begin{verbatim}
import Prelude hiding ( ($), succ, pred ) -- reclaim those names

class Funktion ft where
  apply :: (ft t u) -> t->u
  ($)   :: (ft t u) -> t->u
  f $ x = apply f x
  compose :: (ft t u) -> (ft s t) -> (ft s u)

instance Funktion (->) where
  apply   f x = f x  ...

data Fn t u = Fn (t->u) ...
...
class (Funktion ft) => HasInverse ft where
  inv :: ft u v -> ft v u

data FnI t u = FnI (t->u) (FnI u t)  -- say

instance Funktion FnI where
  apply   (FnI f _) = f
  compose (FnI f fi) (FnI g gi) = FnI (compose f g) (compose gi fi)
...
class (Funktion ft) => Differentiable ft where
  d_dx :: (Floating t) => ft t t -> ft t t
  plus :: (Floating t) => ft t t -> ft t t -> ft t t
  neg  :: (Floating t) => ft t t -> ft t t  -- ... etc.

instance Differentiable (->) where
  d_dx f x = (f$(x+0.001)-f$(x-0.001))/0.002 -- ... etc

data FnD u v = (Floating u, Floating v, u~v) => FnD (u->v) (FnD u v) --GADTs

instance Funktion FnD where
  apply (FnD f _) = f
  compose (FnD f df) (FnD g dg) =
    FnD (\x -> f (g x))
        (let fg' = (FnD (\x -> (df $ (g x))*(dg $ x)) (d_dx fg')) in fg')

instance Differentiable FnD where
  d_dx (FnD f df) = FnD (apply df) (d_dx df)  -- ...etc.
...
data FnID u v = (Floating u, Floating v, u~v) =>
  FnID (u->v) (FnID u v) (FnID u v) -- i.e., FnID f inv(f) f'
...
\end{verbatim}
\caption{Possible class Funktion in Haskell}
\label{fig:Haskell}
\end{figure*}

\section{What Subclasses?}
\label{sec:Subclasses}

The subclasses of functions used in section \ref{sec:Trials}
are plausible but they were principally chosen to provide a small, simple,
fixed target for experiments in different programming languages.
Surely there is no one right answer to the questions of
what is the best collection of subclasses of class \texttt{Funktion} and
where should any such subclasses be defined --
as built-in to the language,
in the standard library or
in some specialised library --
but the questions are worth considering.

There are various reasons one might create a subclass of \texttt{Funktion}.
One reason
is that some subset of functions has special properties
that allow better ways to implement some standard operations.
For example, a permutation (array) over $[0,n)$ is a natural representation
of a permutation function (in fact a \texttt{HasInverse})
in the symmetric group, $S_n$, and
the composition of two such has the same kind of representation
which can be computed easily and is more efficient to use
in that form than \texttt{Funktion}'s default version of \texttt{compose}.
In a like vein,
a matrix can represent a linear transformation of $\textbf{R}^n$ and
similar remarks apply to the composition of two of them.

Another reason to create a subclass
is that some subset of functions has extra specific properties
that are widely useful.
For example, subclasses could be created for differentiable functions, say
\texttt{R\_2\_R}
for those of type $\textbf{R} \rightarrow \textbf{R}$
having method \texttt{d\_dx} as seen already,
\texttt{Rn\_2\_R}
for those of type $\textbf{R}^n \rightarrow \textbf{R}$
having $\nabla$ (grad),
\texttt{Rn\_2\_Rm}
for those of type $\textbf{R}^n \rightarrow \textbf{R}^m$
having Jacobian \textbf{J}, and
\texttt{Rn\_2\_Rn}
for those of type $\textbf{R}^n \rightarrow \textbf{R}^n$
the Jacobian being square and therefore having a determinant.
These properties are relevant to optimisation algorithms.
Incidentally, the \textbf{C} 1-D root finding routines in
the GNU Scientific Library \cite{gsl}
need to be given a procedure, \texttt{fdf},
that calculates both $f(x)$ and $f'(x)$ at the same time;
a default \texttt{fdf} method to do this
could be provided in subclass Differentiable
by using \texttt{apply} and \texttt{d\_dx},
the default being overridden if desired on efficiency grounds.
Similar remarks apply to the multidimensional case.

Note that a mathematical function $f : X \rightarrow Y$ has an inverse
$f^{-1} : Y \rightarrow X$ if and only if $f$ is a bijection.
There are weaker kinds of inverse:
a  left-inverse $f^{-1}_l$ if and only if
$f^{-1}_l \circ f$ is the identity on $X$ and
a right-inverse $f^{-1}_r$ if and only if
$f \circ f^{-1}_r$ is the identity on $Y$,
for example $arcsin : [-1,1] \rightarrow [-\frac{\pi}{2},\frac{\pi}{2}]$
is a right-inverse of $sin : (-\infty,\infty) \rightarrow [-1,1]$.
It may or may not be worth having subclasses to cover these distinctions, or
alternatively \texttt{HasInverse} might just accept an
$f^{-1}_l$ or an $f^{-1}_r$ as ``inverse''.

It might be worth having one or more subclasses of functions
producing numerical results (\texttt{Num}s in Haskell terms)
because $+, -, *$ and $/$ are applicable to them
as in $(f+g)(x) = f(x)+g(x)$ and $(f+g)' = f'+g'$.

Defining integrals has much in common with defining derivatives --
a default method can be implemented by Simpson's rule, say,
to be overridden where there is a closed form.
Like a derivative,
an integral \texttt{Funktion} (as opposed to its result) is to be
computed lazily in general because an integral will itself be integrable.

Looking at the standard libraries of
Javascript \cite{jsLib}, Java \cite{javaLib},
C++ \cite{cppLib} and Python \cite{pyLib}
there is a high degree of commonality
their largely being populated with functions of types such as
$\textbf{R} \rightarrow \textbf{Z}$ (round, floor, ceil, ...),
$\textbf{R} \rightarrow \textbf{R}$ (abs, sin, asin, cos, exp, log, ...)
and
$\textbf{R}^2 \rightarrow \textbf{R}$ (pow, min, max, ...),
\textit{suggesting} that anything more exotic
might be reserved for specialised libraries.

\section{Conclusion}
\label{Conclusion}

It is not surprising that
a programmer can define a \texttt{class Funktion}
with methods \texttt{apply} and \texttt{compose},
and subclasses,
in object-oriented languages such as Javascript, Java, C++ and Python,
and in Haskell.
However, this leads one to wonder if such a class,
probably with the name ``Function'',
could be fully incorporated into such languages
with instances being
those functions coming \textit{as standard} with those languages and not
just external libraries and user-defined functions added later.
Subclasses of \texttt{Funktion} can specify
special kinds of function that have extra properties --
inverses and derivatives in the above (sec.\ref{sec:Trials})
but more are very much possible (sec.\ref{sec:Subclasses}).

Adopting class Function, and subclasses, would bring
a programming language closer to mathematical usage of functions,
increasing expressive power -- and would allow the compiler or interpreter
to detect more kinds of programming error.

C++ (sec.\ref{sec:Cpp}) and Python (sec.\ref{sec:Python})
allow programmers to overload
implicit application ($(\,)$)
so that a \texttt{Funktion}, \texttt{f},
can be called by writing \texttt{f(x)},
circumventing having to write \texttt{f.apply(x)},
and similarly for \texttt{f.inv(y)} and \texttt{f.d\_dx()(x)} if \texttt{f}
has an inverse and a derivative.

In general a \texttt{Funktion},
particularly the result of an operator on a \texttt{Funktion},
does have to be allocated in the heap.
If this is felt to be too inefficient for a simple use of
a function such as \texttt{math.abs}, it would not be difficult
to automatically provide a \texttt{class Funktion} wrapper
for the basic function only when needed.
The presence of a garbage collector in a programming language
also makes life easier.

Lest it be said that
\texttt{class Funktion} and subclasses are just syntactic sugar,
the same could be said of functional programing languages and
object-oriented programming languages themselves --
after all, everything computable can be programmed in assembly code,
it is just a lot easier and more reliable not to do so.

\section*{Acknowledgements}
Thanks to
Arun Konagurthu for discussions about C++ and to
Nathan Hurst for discussions about C++ and Python.


\bibliographystyle{plainurl}   


\bibliography{paper}  

\begin{thebibliography}{10}

\bibitem{cppLib}
{C}++ math, 2024.
\newblock Accessed May 2024.
\newblock URL: \url{https://www.w3schools.com/cpp/cpp_ref_math.asp}.

\bibitem{javaLib}
Java, lang.{M}ath, 2024.
\newblock Accessed May 2024.
\newblock URL:
  \url{https://docs.oracle.com/javase/8/docs/api/java/lang/Math.html}.

\bibitem{jsLib}
Javascript, {M}ath object, 2024.
\newblock Accessed May 2024.
\newblock URL: \url{https://www.w3schools.com/jsref/jsref_obj_math.asp}.

\bibitem{operators}
List of mathematic operators, 2024.
\newblock Accessed May 2024.
\newblock URL:
  \url{https://en.wikipedia.org/wiki/List_of_mathematic_operators}.

\bibitem{pyLib}
Python's math -- mathematical functions, 2024.
\newblock Accessed May 2024.
\newblock URL: \url{https://docs.python.org/3/library/math.html}.

\bibitem{pytoolz}
pytoolz/toolz -- a functional standard library for python, 2024.
\newblock Accessed January 2025.
\newblock URL: \url{https://github.com/pytoolz/toolz}.

\bibitem{gsl}
{GNU} {Scientific} {L}ibrary, 2025.
\newblock Accessed January 2025.
\newblock URL: \url{https://www.gnu.org/software/gsl/doc/html/}.

\bibitem{All18}
Lloyd Allison.
\newblock {\em Coding Ockham's Razor}.
\newblock Springer, 2018.
\newblock \href {https://doi.org/10.1007/978-3-319-76433-7}
  {\path{doi:10.1007/978-3-319-76433-7}}.

\bibitem{All22}
Lloyd Allison.
\newblock Subclasses of class function used to implement transformations of
  statistical models.
\newblock {\em ArXiv}, pages 1--7, July 2022.
\newblock \href {https://doi.org/10.48550/arXiv.2207.04218}
  {\path{doi:10.48550/arXiv.2207.04218}}.

\bibitem{Bac78}
John Backus.
\newblock Can programming be liberated from the von {N}eumann style?: A
  functional style and its algebra of programs.
\newblock {\em Communications of the ACM}, 21(8):613--641, 1978.
\newblock \href {https://doi.org/10.1145/359576.359579}
  {\path{doi:10.1145/359576.359579}}.

\bibitem{Bac81}
John Backus.
\newblock Function level programs as mathematical objects.
\newblock In {\em Proc. ACM Conf. on Functional Languages and Computer
  Architecture}, pages 1--10. ACM, October 1981.
\newblock \href {https://doi.org/10.1145/800223.806757}
  {\path{doi:10.1145/800223.806757}}.

\bibitem{Chu41}
Alonzo Church.
\newblock {\em The Calculi of Lambda Conversion}.
\newblock Number~6 in Annals of Mathematical Studies. Princeton University
  Press, 1941.
\newblock URL: \url{https://press.princeton.edu/titles/2390.html}.

\bibitem{Python}
Fred~L. Drake~Jr.
\newblock {\em Python Language Reference Manual}.
\newblock Network Theory Ltd, 2003.
\newblock Also see \href{http://www.python.org/}{www.python.org}.

\bibitem{JS}
David Flanagan.
\newblock {\em JavaScript -- The Definitive Guide}.
\newblock O'Reilly, 2020.
\newblock Also see
  \href{http://developer.mozilla.org/en-US/docs/Web/javascript}{developer.mozilla.org/en-US/docs/Web/javascript}.

\bibitem{Java}
James Gosling, Bill Joy, and Guy Steele.
\newblock {\em The Java Language Specification}.
\newblock Addison Wesley, 1997.
\newblock Also see \href{http://www.java.com}{www.java.com}.

\bibitem{APL}
Kenneth~E. Iverson.
\newblock {\em A Programming Language}.
\newblock Wiley, 1962.
\newblock Also see J.

\bibitem{J}
Kenneth~E. Iverson.
\newblock {\em Programming in J}.
\newblock Iverson Software Inc., Toronto, 1992.
\newblock Also see \href{http://www.jsoftware.com}{www.jsoftware.com}.

\bibitem{Lan66}
Peter~J. Landin.
\newblock The next 700 programming languages.
\newblock {\em Communications of the ACM}, 9(3):157--166, 1966.
\newblock \href {https://doi.org/10.1145/365230.365257}
  {\path{doi:10.1145/365230.365257}}.

\bibitem{Haskell2010}
Simon Marlow, editor.
\newblock {\em Haskell 2010 Language Report}.
\newblock 2010.
\newblock URL: \url{http://www.haskell.org/onlinereport/haskell2010/}.

\bibitem{HaskellPredef}
Simon Marlow, editor.
\newblock {\em Haskell 2010 {L}anguage {R}eport}, chapter 6, {P}redefined
  {Ty}pes and {C}lasses, figure\ 6.1.
\newblock 2010.
\newblock URL:
  \url{https://www.haskell.org/onlinereport/haskell2010/haskellch6.html}.

\bibitem{Haskell98}
Simon Peyton~Jones, editor.
\newblock {\em Haskell 98 Language and Libraries, the Revised Report}.
\newblock Cambridge University Press, 2003.
\newblock Also see \href{http://www.haskell.org/}{www.haskell.org}.

\bibitem{Str67}
Christopher Strachey.
\newblock Fundamental concepts of programming languages.
\newblock In {\em International Summer School in Computer Programming}, pages
  1--39, Copenhagen, August 1967.
\newblock \href {https://doi.org/10.1023/A:1010000313106}
  {\path{doi:10.1023/A:1010000313106}}.

\bibitem{Cpp}
Bjarne Stroustrop.
\newblock {\em The C++ Programming Language 4th ed.}
\newblock Addison Wesley, 2014.
\newblock Also see
  \href{http://en.wikipedia.org/wiki/C++}{en.wikipedia.org/wiki/C++}.

\bibitem{A68}
Adriaan van Wijngaarden, Barry~J. Mailloux, John E.~L. Peck, Cornelis H.~A.
  Koster, Michel Sintzoff, Charles~H. Lindsey, Lambert G. L.~T. Meertens, and
  Richard~G. Fisker, editors.
\newblock {\em Revised Report on the Algorithmic Language Algol~68}.
\newblock Springer, 1976.
\newblock \href {https://doi.org/10.1145/954652.1781176}
  {\path{doi:10.1145/954652.1781176}}.

\bibitem{H2002}
Various.
\newblock {c}lass {F}unction?, October 2002.
\newblock Discussion in the {H}askell mailing list (29~Oct. 2002); see extract
  at
  \href{https://www.cantab.net/users/mmlist/ll/BIB/Local/2002Haskell.html}{www.cantab.net/users/mmlist/ll/BIB/Local/2002Haskell.html}$\leftarrow$(click).

\end{thebibliography}

\end{document}